\documentclass{article}
\usepackage[T1]{fontenc} 
\usepackage[utf8]{inputenc} 
\usepackage[lbd]{ismir}
\usepackage{amsmath,cite,url}
\usepackage{graphicx}
\usepackage{color}
\usepackage{booktabs}
\usepackage{graphicx}
\usepackage{cuted}
\usepackage{capt-of}
\usepackage[bookmarks=false]{hyperref}
\usepackage{multirow}
\usepackage{enumitem}

\usepackage{booktabs}
\usepackage{graphicx}
\usepackage{dblfloatfix}

\usepackage{xcolor}

\newcommand{\result}[2]{%
  #1\,{\scriptsize\color{black!45}[#2]}%
}

\title{MuSP-Bench: Advanced Multimodal Benchmarking of Music Understanding Across Score and Performance}

\multauthor
{Milan Liessens Dujardin$^1$$^2$ \hspace{1cm} Song-Ze Yu$^2$ \hspace{1cm} Kevin Miao$^1$}
{
$^1$ Bryel Labs\hspace{1cm} $^2$ UC Berkeley \\
{\ttfamily\small milan.ld@berkeley.edu}
\quad
{\ttfamily\small vaclis@berkeley.edu}
\quad
{\ttfamily\small kevin@bryel.ai}
}

\def\authorname{M. Liessens Dujardin, S. Yu, and K. Miao}

\begin{document}

\maketitle

\begin{abstract}
Musicians commonly communicate musical intent through scores and performances. To investigate whether models can reason across both modalities, we introduce \mbox{\textbf{MuSP-Bench}}, a human-authored benchmark of 490 questions that assess understanding across \textbf{Mu}sical \textbf{S}cores and \textbf{P}erformances, both individually and in combination. The benchmark spans interpretive and long-horizon reasoning across classical piano and orchestral works, represented through ABC notation, score images, MIDI, and audio.
Our evaluation of frontier multimodal large language models reveals substantial difficulties in understanding musical scores and even greater challenges in reasoning about audio. The benchmark is available at \url{https://musp.vaclis.net/}.
\end{abstract}

\section{Introduction}\label{sec:introduction}

Musicians commonly express musical ideas through both scores and performances. Scores encode musical intent through pitch, rhythm, dynamics, articulation, and structure, while performances realize that intent in sound. Musicians continually reason both within and across these modalities when playing, interpreting, imagining, and analyzing music. Artificial musical intelligence capable of supporting real-world applications in education and artistic practice should therefore do the same across these modalities and their common representations: ABC notation, score images, MIDI, and audio.

Existing benchmarks mainly assess these modalities in isolation. For instance, MusicTheoryBench \cite{yuanChatMusicianUnderstandingGenerating2024b} evaluates understanding of mostly short ABC excerpts; MMMU's music subset \cite{yueMMMUMassiveMultiDiscipline2024a} and WildScore \cite{mundadaWildScoreBenchmarkingMLLMs2025a} primarily focus on single-image scores; ZIQI-Eval \cite{liMusicMaestroMusically2024b} mostly probes factual knowledge, though it includes an ABC continuation task; these benchmarks predominantly rely on multiple-choice evaluation. ABC-Eval \cite{zhaoABCEvalBenchmarkingLarge2025b} is limited to ABC; MSU-Bench \cite{daiMusicalScoreUnderstanding2026b} evaluates complete scores, but predominantly through short-horizon or binary questions. MusiXQA \cite{chenMusiXQAAdvancingVisual2025b} and SSMR-Bench \cite{wangAIMusicianSynthesizing2025e} generate synthetic scores and questions—while useful for symbolic music understanding, they provide limited coverage of higher-level musical interpretation. 

On the other hand, benchmarks like HumMusQA \cite{weckHumMusQAHumanwrittenMusic2026b}, MUSE \cite{carone2025musebenchmarkprobingmusic}, CMI-Bench \cite{maCMIBENCHCOMPREHENSIVEBENCHMARK2025a}, MusTBench \cite{kwonMusTBENCHBenchmarkingAdvancing2026c}, and PitchBench \cite{pitchbench} assess musical perception from audio, including pitch, genre, melody, and instrumentation, as well as temporal grounding. However, none of these benchmarks assess understanding of performance and interpretative choices in relation to the score. MuseBench \cite{zhaoMuseAgent1InteractiveGrounded2026b} tests reasoning over score and audio, but its audio-input tasks focus on lower-level performance attributes, such as key accuracy, completeness, tempo stability, and speed through true/false questions, without addressing voicing, phrasing, technique, and expression. 

Overall, existing benchmarks rarely evaluate musician-like reasoning across score and performance: locating a performance’s climax in the score, analyzing phrasing, or interpreting a composer's intent. We therefore introduce a multimodal benchmark that combines fine-grained music understanding with long-horizon reasoning and real-world inquiries across score and performance.

\section{Benchmark}
\label{sec:benchmark}

MuSP-Bench comprises 490 questions across 18 complete classical piano pieces or movements and 6 orchestral excerpts, assessing understanding of scores, performances, and both jointly. Of these, 460 are open-ended within a predefined answer format, such as a pitch, bar, timestamp, or annotation, allowing multiple correct answers to reflect musical ambiguity; 30 provide multiple options when a precise format could not be specified. Piano scores come from (n)ASAP \cite{peterAutomaticNoteLevelScoretoPerformance2023d}, cleaned using consensus from PianoCoRe performance MIDI \cite{borovikPianoCoReCombinedRefinedb}; orchestral scores and audio come from BSED \cite{berendesBeethovenSymphonyExcerptb}. The benchmark was authored by two musicians with over 20 years of experience.

Each question is annotated along four dimensions. \textbf{Modality} indicates whether the minimum required information comes from the \textit{score} (S), \textit{performance} (P), \textit{both} jointly (S\&P), or \textit{either} (S/P). \textbf{Horizon} indicates the minimum evidence span: \textit{short} (1--2 bars or less than $\sim$5 s; 19\% of the questions), \textit{medium} (3--16 bars or approximately 5--30 s; 25\%), \textit{long} (larger or noncontiguous; 36\%), or \textit{any} (any part can be used as evidence; 20\%). Pieces average approximately 10 score pages (range: 1--31) and 5 minutes (range: 0:17--12:29). \textbf{Content} follows a hierarchy from \textit{pitch}, through \textit{temporal organization}, \textit{orchestration}, and \textit{performance realization}, to \textit{melodic}, \textit{harmonic}, \textit{formal}, \textit{interpretive}, and \textit{contextual} understanding. \textbf{Operation} covers identification, localization, quantification, comparison, ordering, inference, transcription, explanation, segmentation, or summarization.\footnote{\label{fn:hf-details}Additional statistics, MIDI formats, and ablation results: \url{https://huggingface.co/datasets/bryel-labs/MuSP-Bench}.}


\subsection{Question Overview}

MuSP-Bench covers score analysis (S), performance understanding (P), score--performance integration (S\&P), and questions answerable from either modality (S/P). It targets harmony, form, phrasing, voicing, expression, motifs, themes, articulation, pedaling, texture, style, tonality, composer, and work identification. Examples include:
\begin{itemize}[
    labelindent=0pt,
    leftmargin=*,
    itemsep=1pt,
    topsep=2pt,
    parsep=0pt,
    partopsep=0pt
]
    \item S\&P: Consider the theme starting at 2:51. In which bar does the performer reach the subverted, silent climax of this theme, characterized by a significant decrease in tempo and loudness?
    \item S/P: What note creates the effect of a relentless pulse in the beginning of the recapitulation?
    \item P: How many F3 bass notes are played truly staccato in the performance before 0:08 (i.e., not elongated in any way)?
\end{itemize}

\section{Experiments}
\label{sec:experiments}

We evaluate four frontier multimodal foundation models on their supported input modalities: GPT-5.6 Sol \cite{openAIGPT56SystemCard2026} on text and score images, AF-Next-Instruct \cite{ghoshAudioFlamingoNext2026b} on audio, and Muse Spark 1.2 \cite{metaMuseSpark2026} and Qwen3.5-Omni-Plus \cite{teamQwen35OmniTechnicalReport2026b} on all three formats. For S and S/P questions, we use ABC notation and PDF scores converted to images; for P and S/P: audio and MIDI rendered as text using scientific pitch notation (referred to as MIDI-as-text v1); for S\&P: score images with audio (image+audio) as well as ABC with audio (ABC+audio) or ABC with MIDI (ABC+MIDI). We remove work metadata from the headers.
A language-only baseline measures how far models can answer using only composer and title metadata; see \textit{Metadata} in Table \ref{tab:main-results}. As an ablation, we assess representation sensitivity by comparing MIDI-as-text v1 with an alternative serialization (v2) that uses numeric pitch values.\textsuperscript{\ref{fn:hf-details}} Reported scores use task-dependent deterministic answer normalization.

\section{Results}

On S questions, GPT-5.6 Sol and Muse Spark 1.2 perform best with ABC notation, while on P questions, they perform best with MIDI-as-text, highlighting their advantage with textual representations. Qwen3.5-Omni-Plus performs similarly across formats for S and P. 

On S\&P questions, Muse Spark's accuracy drops by 25.0 percentage points when replacing ABC+MIDI with either image+audio or ABC+audio, suggesting that reasoning jointly over score and audio remains challenging.

The S/P subsets reveal task-dependent modality strengths. Textual inputs perform best on general analytical questions, with GPT achieving the highest accuracy using MIDI-as-text or ABC. Audio and images perform better for global attributes: for tonality/style, Qwen with audio outperforms GPT with textual inputs, while Muse Spark leads with images. Images also benefit composer/title identification, where Muse Spark performs best.

Sensitivity to textual MIDI representation varies across models, with accuracy dropping by 24.4, 7.6, and 0.4 percentage points for GPT, Qwen, and Muse Spark, respectively.\textsuperscript{\ref{fn:hf-details}}

In summary, on textual inputs, GPT is the strongest; visual and audio inputs offer particular benefits for stylistic and contextual understanding. 


\providecommand{\result}[2]{}
\renewcommand{\result}[2]{#1\,{\scriptsize\textcolor{black!45}{[#2]}}}
\providecommand{\notapp}{}
\renewcommand{\notapp}{\textcolor{black!45}{\textemdash}}
\providecommand{\notrun}{}
\renewcommand{\notrun}{\textcolor{black!45}{\textnormal{---}}}
\par\medskip
\noindent
\begin{minipage}{\columnwidth}
\centering
\captionof{table}{Model performance, measured by accuracy (\%). Bold: row best; underline: model’s best format per subset (excluding metadata control condition); —: inputs not natively supported. The weighted max selects the highest non-metadata accuracy within each subset and weights these scores by the respective question counts.}
\label{tab:main-results}
\vspace{2pt}
\tiny
\setlength{\tabcolsep}{2pt}
\renewcommand{\arraystretch}{1.08}
\newlength{\muspinputwidth}
\settowidth{\muspinputwidth}{MIDI-as-text (ablation)}
\resizebox{\columnwidth}{!}{%
\begin{tabular}{@{}llrcccc@{}}
\toprule
\textbf{Subset} & \textbf{Input} & \textbf{\#Q} &
\shortstack{\textbf{GPT-5.6}\\\textbf{Sol}} &
\shortstack{\textbf{Qwen3.5-}\\\textbf{Omni-Plus}} &
\shortstack{\textbf{Muse Spark}\\\textbf{1.2}} &
\shortstack{\textbf{AF-Next-}\\\textbf{Instruct}} \\
\specialrule{0.10em}{0pt}{1pt}
\multirow{3}{*}{S} & Metadata & 180 & \textbf{5.0} & 2.2 & 4.4 & \notrun \\
 & ABC & 180 & \textbf{\underline{51.7}} & 11.7 & \underline{31.7} & \notrun \\
 & Image & 180 & \textbf{27.2} & \underline{12.2} & 15.0 & \notapp \\
\specialrule{0.10em}{2pt}{1pt}
\multirow{3}{*}{P} & Metadata & 106 & \textbf{6.6} & 4.7 & \textbf{6.6} & 3.8 \\
 & MIDI-as-text v1 & 106 & \textbf{\underline{66.0}} & \underline{16.0} & \underline{38.7} & \notrun \\
 & Audio & 106 & \notapp & \textbf{14.2} & 3.8 & \underline{0.9} \\
\specialrule{0.10em}{2pt}{1pt}
\multirow{4}{*}{S\&P} & Metadata & 72 & \textbf{8.3} & 4.2 & 4.2 & \notrun \\
 & ABC+MIDI & 72 & \textbf{\underline{45.8}} & \underline{5.6} & \underline{31.9} & \notrun \\
 & Image+audio & 72 & \notapp & \underline{5.6} & \textbf{6.9} & \notapp \\
 & ABC+audio & 72 & \notapp & 4.2 & \textbf{6.9} & \notrun \\
\specialrule{0.10em}{2pt}{1pt}
\multirow{5}{*}{\shortstack[l]{S/P\\(General)}} & Metadata & 36 & \textbf{13.9} & 5.6 & 8.3 & 2.8 \\
 & ABC & 36 & \textbf{\underline{44.4}} & \underline{16.7} & 33.3 & \notrun \\
 & MIDI-as-text v1 & 36 & \textbf{\underline{44.4}} & 13.9 & \underline{36.1} & \notrun \\
 & Image & 36 & \textbf{19.4} & \underline{16.7} & 11.1 & \notapp \\
 & Audio & 36 & \notapp & \textbf{\underline{16.7}} & 11.1 & \underline{0.0} \\
\specialrule{0.10em}{2pt}{1pt}
\multirow{4}{*}{\shortstack[l]{S/P\\(Tonality/style)}} & ABC & 48 & \textbf{\underline{77.1}} & 50.0 & 68.8 & \notrun \\
 & MIDI-as-text v1 & 48 & \textbf{72.9} & 41.7 & 52.1 & \notrun \\
 & Image & 48 & 68.8 & 52.1 & \textbf{\underline{91.7}} & \notapp \\
 & Audio & 48 & \notapp & \textbf{\underline{79.2}} & 20.8 & \underline{22.9} \\
\specialrule{0.10em}{2pt}{1pt}
\multirow{4}{*}{\shortstack[l]{S/P\\(Composer/title)}} & ABC & 48 & 35.4 & 14.6 & \textbf{54.2} & \notrun \\
 & MIDI-as-text v1 & 48 & \textbf{27.1} & 0.0 & 25.0 & \notrun \\
 & Image & 48 & \underline{58.3} & 16.7 & \textbf{\underline{79.2}} & \notapp \\
 & Audio & 48 & \notapp & \textbf{\underline{54.2}} & 2.1 & \underline{0.0} \\
\specialrule{0.10em}{2pt}{1pt}
\multirow{9}{*}{\shortstack[l]{Per-model\\format\\comparison}} & Metadata & 394 & \textbf{6.9} & 3.6 & 5.3 & \result{3.5}{n=142} \\
 & ABC & 312 & \textbf{52.2} & 18.6 & 41.0 & \notrun \\
 & MIDI-as-text v1 & 238 & \textbf{56.3} & 17.6 & 38.2 & \notrun \\
 & \makebox[\muspinputwidth][l]{MIDI-as-text v2} & 238 & 31.9 & 10.1 & \textbf{37.8} & \notrun \\
 & Image & 312 & \textbf{37.5} & 19.6 & 36.2 & \notapp \\
 & Audio & 238 & \notapp & \textbf{35.7} & 8.0 & 5.0 \\
 & ABC+MIDI & 72 & \textbf{45.8} & 5.6 & 31.9 & \notrun \\
 & Image+audio & 72 & \notapp & 5.6 & \textbf{6.9} & \notapp \\
 & ABC+audio & 72 & \notapp & 4.2 & \textbf{6.9} & \notrun \\
\specialrule{0.14em}{3pt}{2pt}
\textbf{Aggregate} & \textbf{Weighted max} & \textbf{490} & \textbf{56.5} & 23.1 & 44.1 & \notrun \\
\bottomrule
\end{tabular}%
}
\end{minipage}
\par\medskip

\section{Conclusion}
\label{sec:conclusion}



We introduce MuSP-Bench, a human-authored benchmark for interpretive and analytical reasoning across musical scores and performances, spanning both short- and long-horizon tasks. Our evaluation reveals task-dependent modality strengths: on analytical questions, models generally perform best with textual representations, while images and audio offer advantages for global stylistic and contextual understanding. However, performance audio and image-based scores remain challenging overall. These findings position MuSP-Bench as a foundation for tracking progress toward true multimodal music understanding.



\bibliography{ISMIRtemplate}

\end{document}